\documentclass[aps,prd,reprint,nofootinbib]{revtex4-1}
\usepackage{amsmath,amssymb}
\usepackage[dvips]{hyperref,graphicx}

\newcommand{\bfx}{\textbf{x}}
\newcommand{\bfy}{\textbf{y}}
\newcommand{\bfz}{\textbf{z}}
\newcommand{\bfp}{\textbf{p}}
\newcommand{\bfq}{\textbf{q}}

\newcommand{\bfgamma}{{\boldsymbol \gamma}}
\newcommand{\bfnabla}{{\boldsymbol \nabla}}

\newcommand{\fslash}[1]{{\ooalign{\hfil/\hfil\crcr$#1$}}}

\begin{document}
\title{Dark Energy from Eternal Pair-production of Fermions}
\author{Jiro \surname{Hashiba}}

\begin{abstract}
We study a toy model in which Majorana or Dirac fermions behave as the source for a small vacuum energy in the present Universe. In the model, a self-interacting scalar boson coupled with fermions induces attractive and repulsive interactions between the fermions simultaneously. These interactions allow for the existence of a metastable state with positive energy density comprised of fermions degenerate inside a Fermi surface. The energy density of the metastable state remains constant as the Universe expands. This is because pair-productions of fermions from the vacuum continuously take place at no energy cost and keep supplying fermions uniformly to the Universe. The observed vacuum energy density $\sim 10^{-47}~\textrm{GeV}^4$ is reproduced for the fermion and scalar mass of the order $10^{-3}~\textrm{eV}$.
\end{abstract}

\maketitle

\section{Introduction}

In 1998, a couple of independent research groups announced their groundbreaking discoveries in particle physics and astrophysics. One was the neutrino oscillation, which showed that neutrinos had mass \cite{kamiokande}. The other was the accelerating expansion of the Universe \cite{riess-perlmutter}. A non-vanishing cosmological constant is the simplest explanation of this observation. Nevertheless, it cannot be ruled out that the accelerating expansion is caused by a form of dynamically generated energy, which is now dubbed \textit{dark energy} \cite{copeland-sami-tsujikawa}. The year of announcement was not the only coincidence of these discoveries. If the dark energy arises from a zero-point energy inherent in quantum field theory, its density is expected to take a value $\sim M_P^4$, where $M_P$ is the Planck scale. However, the quartic root of the observed dark energy density is of the order $10^{-3}~\textrm{eV}$ \cite{planck2018-6}, which is not far from the upper bound of active neutrino mass $\sim 2~\textrm{eV}$ \cite{pdg2016}. Motivated by this mass scale similarity, some researchers have attempted to explain the connection between the dark energy and neutrinos. Among such researches are models based on mass varying neutrino (MaVaN) \cite{hung,fardon-nelson-weiner,peccei}, neutrino condensate \cite{caldi-chodos,bhatt-desai-ma-rajasekaran-sarkar}, and the vacuum condensate of neutrino mixing \cite{capolupo-capozziello-vitiello}. In addition, the author of \cite{lake} has recently presented a qualitative argument that, if the space is filled with ``dark energy particles'' with mass $\sim 10^{-3}~\textrm{eV}$, the pair-production of such particles leads to an accelerated expansion of the Universe. In this article, we propose a specific model in which the vacuum energy originates from non-trivial dynamics of fermions induced by a scalar field. It is shown that, similarly to the existing literature, the energy scale of the dark energy can be explained if the fermion has mass comparable to that of active neutrinos. However, as shown in the article, the fermion in the model cannot be identified with active neutrinos, but possibly with a fermion which has not yet been discovered.

This article is organized as follows. Section \ref{section-model} is the main part of this article, where we introduce the model Lagrangian and present the thermal history of the particles in the model. In section \ref{section-metastable}, we show the existence of a metastable state with a positive energy density which is identified with the dark energy. The present Universe has been trapped in this metastable state. The concept ``eternal pair-production,'' which is responsible for the stability of the dark energy, is explained in section \ref{section-pair-production}. In section \ref{section-fate}, we estimate the probability that the metastable state decays into the ground state with vanishing energy density. This estimation is crucial because we may live in a Universe without dark energy unless the probability is extremely small. While sections \ref{section-model} through \ref{section-fate} are devoted to the discussion of Majorana fermion case, we briefly demonstrate that our model works in case of Dirac fermion as well in section \ref{section-dirac}. In the last section, we summarize our results and discuss some remaining issues.

\section{The model}
\label{section-model}

We assume a homogeneous, isotropic, and flat Universe described by the Robertson-Walker metric. The scale factor $a$, total energy density $\rho_T$, and total pressure $p_T$ of the Universe evolve according to the following three equations \cite{kolb-turner}:
\begin{equation}
\label{friedmann}
\begin{aligned}
 \left(\frac{\dot{a}}{a}\right)^2 &= \frac{\rho_T}{3M_P^2}, \\
 d(\rho_T a^3) &= -p_Td(a^3), \\
 p_T &= w_T\rho_T,
\end{aligned}
\end{equation}
where $w_T$ is a parameter that is not necessarily a constant. In (\ref{friedmann}), the first, second, and third lines represent the Friedmann equation without the cosmological term, the energy conservation law, and the equation of state, respectively. We assume that there are two sectors in the matter content that contribute to $\rho_T$ and $p_T$. One is the dark energy (DE) sector which accounts for the energy that dominates the present Universe. The DE sector is introduced in the next subsection. The other is whatever else including the standard model (SM) particles and dark matter. In this article, this sector is called the SM sector. We postulate that the coupling between the SM sector and DE sector is highly suppressed, because the particles in the DE sector should have already been detected otherwise.

This section is divided into three subsections. In the first (resp. second) subsection, we focus only on the DE sector, assuming that the temperature is higher (resp. lower) than the characteristic scale of the DE sector. In the last subsection, we explain that the DE sector does not spoil the good prediction of the standard cosmology in the early Universe, and begins dominating the Universe only recently.

\subsection{High temperature phase}

The DE sector consists of a Majorana fermion $\psi$ and a real scalar $\phi$. We mainly consider the case where $\psi$ is a Majorana spinor in this article, but our model also works for the case of Dirac fermion. The Dirac fermion case will be briefly studied in section \ref{section-dirac}. The fermion and the scalar interact with each other through a Yukawa coupling. The Lagrangian of the model is given by
\begin{align}
 \label{lag}
\begin{aligned}
 L &\equiv \int d^3x\left[
              \frac{1}{2}\bar{\psi}\left(i\fslash{\partial}-m\right)\psi
              -\frac{g}{2}\phi\bar{\psi}\psi\right. \\
   & \qquad\qquad\quad +\left.
              \frac{1}{2}\partial_{\mu}\phi\partial^{\mu}\phi-\frac{1}{2}m_{\phi}^2\phi^2
              -\frac{\lambda}{4}\phi^4\right],
\end{aligned}
\end{align}
where $\psi$ satisfies the Majorana condition $\psi=\psi^c\equiv {\cal C}\bar{\psi}^T$, with ${\cal C}\equiv -i\gamma^2\gamma^0$ being the charge conjugation matrix. We do not include the cubic term of $\phi$ in (\ref{lag}) for simplicity, although $L$ has no symmetry that precludes the cubic term. There exist four model parameters: fermion mass $m$, scalar mass $m_{\phi}$, Yukawa coupling $g$, and scalar self-coupling $\lambda>0$.

Let $T$ be the temperature of the DE sector. Note that $T$ is different from the temperature $T_{SM}$ of the SM sector in general, since the DE sector is almost decoupled from the SM sector. As we will see in the next section, the masses of the DE sector have to satisfy $m\sim m_{\phi}\sim 10^{-3}~\textrm{eV}$ in order for the observed dark energy scale to be explained. The Lagrangian (\ref{lag}) well describes the DE sector for $T\gtrsim 10^{-3}~\textrm{eV}$. On the other hand, a low energy effective Lagrangian is more convenient for $T\lesssim 10^{-3}~\textrm{eV}$, as considered in the next subsection.

Throughout this article, we make the following assumption on the system described by the Lagrangian (\ref{lag}).
\begin{description}
 \item[Assumption 1]\mbox{}\\
The Lagrangian (\ref{lag}) implicitly contains the counter terms which cancel out divergent terms depending on some cutoff scale, say, the Planck scale $M_P$.
\end{description}
A few comments on this assumption are in order. A zero-point energy is an example of the divergent terms supposed in Assumption 1. In this article, we assume that the vacuum energy is generated by a combination of two mechanisms. First, some unknown mechanism like supersymmetry fixes the zero-point energy exactly at zero, which otherwise is expected to take a value $\sim M_P^4$. The first mechanism is followed by a second one that lifts the vacuum energy from zero to the observed tiny level of $(10^{-3}~\textrm{eV})^4$. Our goal is to present the second mechanism, while the first one is beyond the scope of this article.

\subsection{Low temperature phase}

In this subsection, we find a low energy effective theory of (\ref{lag}) which is suitable for describing the dark energy. In this subsection and sections from \ref{section-metastable} to \ref{section-fate}, we make the following assumption.
\begin{description}
 \item[Assumption 2]\mbox{}\\
The temperature of the DE sector satisfies $T\ll m\sim m_{\phi}\sim 10^{-3}~\textrm{eV}$ and the entropy $S$ of the DE sector is sufficiently small so that $TS$ is much smaller than the (internal) energy of the metastable state discussed in section \ref{section-metastable}.
\end{description}
When $T\ll 10^{-3}~\textrm{eV}$, the scalar $\phi$ annihilates and only serves as the virtual particles mediating the interaction between fermions $\psi$. On the other hand, in the next section we will show the existence of a metastable state on which fermionic particle excitations have a dispersion relation with no energy gap (see $E_p$ in (\ref{energy-cost}) below). It is thus justified to integrate out only $\phi$ and retain $\psi$ in the Lagrangian $L$.

Let us derive a low energy effective Lagrangian for $\psi$ by integrating out $\phi$. Assuming that $\phi$ is time-independent, we write down the equation of motion of $\phi$ as
\begin{equation}
 \label{saddle-condition}
 \frac{g}{2}\bar{\psi}\psi+(-\bfnabla^2+m_{\phi}^2)\phi+\lambda\phi^3=0.
\end{equation}
By treating the Laplacian $\bfnabla^2$ as if it is just a numerical coefficient, we can formally solve (\ref{saddle-condition}) for $\phi$ as
\begin{equation}
\label{solution}
\begin{aligned}
 \phi &= \phi_0(\bar{\psi}\psi/2), \\
 \phi_0(z)
         &\equiv \left[\sqrt{\left(\frac{-\bfnabla^2+m_{\phi}^2}{3\lambda}\right)^3
                       +\left(\frac{gz}{2\lambda}\right)^2}
                 -\frac{gz}{2\lambda}\right]^{\frac{1}{3}} \\
   &\quad -\left[\sqrt{\left(\frac{-\bfnabla^2+m_{\phi}^2}{3\lambda}\right)^3
                       +\left(\frac{gz}{2\lambda}\right)^2}
                 +\frac{gz}{2\lambda}\right]^{\frac{1}{3}}.
\end{aligned}
\end{equation}
The formula (\ref{solution}) is a closed form expression for the formal power series expansion  of the solution to (\ref{saddle-condition}) in terms of $\bar{\psi}\psi/2$, namely
\begin{equation}
 \label{approximate-solution}
\begin{aligned}
 \phi &= -g\int d^3y \Delta(\bfx-\bfy)\frac{(\bar{\psi}\psi)(\bfy)}{2} \\
         &\quad +\lambda g^3\int d^3y \Delta(\bfx-\bfy)
                 \left[\int d^3z \Delta(\bfy-\bfz)
                                 \frac{(\bar{\psi}\psi)(\bfz)}{2}\right]^3 \\
         &\quad +{\cal O}((\bar{\psi}\psi/2)^5),
\end{aligned}
\end{equation}
where we defined the propagator, or the Yukawa potential, by
\begin{equation}
 \label{yukawa}
 \Delta(\bfx)\equiv\int\frac{d^3 p}{(2\pi)^3}\frac{e^{i\bfp\cdot\bfx}}{|\bfp|^2+m_{\phi}^2}
   = \frac{e^{-m_{\phi}|\bfx|}}{4\pi|\bfx|}.
\end{equation}
Note that $\phi_0(\cdot)$ is a non-local functional of $\bar{\psi}\psi/2$, which depends on spatial coordinates in general. However, we can let $\phi_0(\cdot)$ be an ordinary function by setting $\bfnabla^2=0$ in (\ref{solution}), when $\bar{\psi}\psi/2$ is a constant. By replacing $\phi$ in (\ref{lag}) with $\phi_0$, we obtain an effective Lagrangian for $\psi$,
\begin{equation}
 \label{effective-lag}
\begin{aligned}
 L_{\textrm{eff}}
   &\equiv \int d^3x\left[\frac{1}{2}\bar{\psi}\left(i\fslash{\partial}-m\right)\psi
                          -g\phi_0\left(\frac{\bar{\psi}\psi}{2}\right)
                           \frac{\bar{\psi}\psi}{2}\right. \\
   &\qquad\qquad
           -\frac{1}{2}\phi_0\left(\frac{\bar{\psi}\psi}{2}\right)(-\bfnabla^2+m_{\phi}^2)
            \phi_0\left(\frac{\bar{\psi}\psi}{2}\right) \\
   &\qquad\qquad
            \left.-\frac{\lambda}{4}\phi_0\left(\frac{\bar{\psi}\psi}{2}\right)^4\right].
\end{aligned}
\end{equation}
The effective Lagrangian $L_{\textrm{eff}}$ contains all tree level interactions mediated by $\phi$. The Hamiltonian derived from (\ref{effective-lag}) is given by
\begin{equation}
 \label{effective-hamiltonian}
\begin{aligned}
 \hat{H} &\equiv \int d^3x\left[\frac{1}{2}\bar{\psi}\left(-i\bfgamma\cdot\bfnabla+m\right)\psi
                                +g\phi_0\left(\frac{\bar{\psi}\psi}{2}\right)
                                 \frac{\bar{\psi}\psi}{2}\right. \\
   &\qquad\qquad\quad
           +\frac{1}{2}\phi_0\left(\frac{\bar{\psi}\psi}{2}\right)(-\bfnabla^2+m_{\phi}^2)
            \phi_0\left(\frac{\bar{\psi}\psi}{2}\right) \\
   &\qquad\qquad\quad
            \left.+\frac{\lambda}{4}\phi_0\left(\frac{\bar{\psi}\psi}{2}\right)^4\right].
\end{aligned}
\end{equation}

Let $V$ be the spatial volume of the system. We expand the field $\psi$ by discrete momentum $\bfp$ as
\begin{equation}
\label{psi}
\begin{aligned}
  \psi(\bfx) &= \sum_{\bfp,\sigma}\frac{1}{\sqrt{2\omega_{|\bfp|}V}}
                \left[u(\bfp,\sigma)e^{i\bfp\cdot\bfx}a(\bfp,\sigma)\right. \\
             & \qquad\qquad\qquad\qquad
                      +\left.v(\bfp,\sigma)e^{-i\bfp\cdot\bfx}a^{\dagger}(\bfp,\sigma)\right],
\end{aligned}
\end{equation}
where $\omega_p\equiv\sqrt{p^2+m^2}$ and $a^{\dagger}(\bfp,\sigma)$ (resp. $a(\bfp,\sigma)$) is the creation (resp. annihilation) operator of the fermion with momentum $\bfp$ and spin index $\sigma$
\footnote{
\label{vector-notation}
In the remainder of the article, we often use the vector notation for creation and annihilation operators like $a_{\bfp}^{\dagger}=(a^{\dagger}(\bfp,+),a^{\dagger}(\bfp,-))^T$ to simplify some formulas.
}. The creation and annihilation operators are normalized so that they satisfy the anti-commutation relation $\{a(\bfp,\sigma),a^{\dagger}(\bfq,\sigma')\}=\delta_{\bfp,\bfq}\delta_{\sigma,\sigma'}$. Let ${\cal H}$ be the Fock space constructed by multiplying creation operators on the vacuum $|0\rangle$ which satisfies $a(\bfp,\sigma)|0\rangle=0$. Then, the internal energy $U$ of the system is
\begin{equation}
 \label{internal-energy}
 U = -\frac{\partial}{\partial\beta}\ln\left(\textrm{Tr}_{{\cal H}}e^{-\beta\hat{H}}\right),
\end{equation}
where $\beta\equiv 1/T$ is the inverse temperature. If the temperature $T$ is sufficiently low as stated in Assumption 2, $U$ is dominated by the ground state in ${\cal H}$. In other words, the information about the local minima of $\hat{H}$ is lost, if we sum over all states in (\ref{internal-energy}). However, the dark energy might correspond to a local minimum of $\hat{H}$, not to the global minimum
\footnote{
Here, by the global or local minimum of $\hat{H}$, we mean taking the extrema of $\langle\Psi|\hat{H}|\Psi\rangle$ with respect to $|\Psi\rangle\in{\cal H}$. For instance, the state $|\Psi\rangle=|\Psi_0\rangle$ that minimizes $\langle\Psi|\hat{H}|\Psi\rangle$ is the ground state of $\hat{H}$, as derived from the variational principle.
}
. Therefore, we shall not calculate $U$, but attempt to search for local minima of $\hat{H}$. Since it is impossible to exhaust all states in ${\cal H}$ when looking for the local minima, we restrict ourselves to a subspace ${\cal H}'$ of ${\cal H}$ defined by
\begin{equation}
 \label{hilbert}
 {\cal H}' \equiv
  \left\{\left.|\Lambda\rangle
                =\left[\prod_{|\bfp|\leq\Lambda,\sigma=\pm}a^{\dagger}(\bfp,\sigma)\right]
                 |0\rangle
         \right|\Lambda\in\mathbf{R}_{\geq 0}\right\}.
\end{equation}
As in (\ref{hilbert}), it is natural to consider only the rotationally symmetric states where fermions are degenerate in all states with $|\bfp|\leq\Lambda$ for some positive $\Lambda$ in an isotropic Universe.

To find local minima of $\hat{H}$, we simplify $\hat{H}$ with a mean field approximation. Specifically, we expand $\hat{H}$ by fermion bilinear terms $\bar{\psi}\bfgamma\cdot\bfnabla\psi$ and $\bar{\psi}\psi$ around their expectation values conditional on the state $|\Lambda\rangle$, and discard the second and higher order terms. Let $\rho(\Lambda)V$ be the zeroth order term, where $\rho(\Lambda)$ equals the energy density $\langle\Lambda|\hat{H}|\Lambda\rangle/V$ under the mean field approximation. Then, $\hat{H}$ is expressed as
\begin{equation}
\label{mean-field-approximation}
 \hat{H} \simeq \rho(\Lambda)V + \hat{H}_0 + \hat{H}_I,
\end{equation}
where $\hat{H}_0 + \hat{H}_I$ is the first order term. We explain how to decompose the first order term into $\hat{H}_0$ and $\hat{H}_I$ later.

Let us first calculate the energy density $\rho(\Lambda)$. The expectation value of the kinetic energy density in (\ref{effective-hamiltonian}) is calculated as
\begin{equation}
 \label{kinetic}
\begin{aligned}
 K(\Lambda)
  &\equiv \left\langle\Lambda\left|\frac{1}{2}\bar{\psi}\left(-i\bfgamma\cdot\bfnabla+m\right)\psi
          \right|\Lambda\right\rangle \\
  &= 2\int_{|\bfp|\leq\Lambda} \frac{d^3p}{(2\pi)^3}\omega_{|\bfp|} \\
  &= \frac{1}{8\pi^2}
     \left[\omega_{\Lambda}\Lambda(m^2+2\Lambda^2)
           +m^4\ln\left(\frac{\omega_{\Lambda}-\Lambda}{m}\right)\right].
\end{aligned}
\end{equation}
We also define the fermionic condensate
\begin{equation}
 \label{potential}
\begin{aligned}
 W(\Lambda)
   &\equiv \left\langle\Lambda\left|\frac{\bar{\psi}\psi}{2}\right|\Lambda\right\rangle \\
   &= 2\int_{|\bfp|\leq\Lambda} \frac{d^3p}{(2\pi)^3}\frac{m}{\omega_{|\bfp|}} \\
   &= \frac{m}{2\pi^2}
     \left[\omega_{\Lambda}\Lambda
           +m^2\ln\left(\frac{\omega_{\Lambda}-\Lambda}{m}\right)\right].
\end{aligned}
\end{equation}
Therefore, we obtain
\begin{equation}
 \label{internal-energy-density}
\begin{aligned}
 \rho(\Lambda) &= K(\Lambda) + P(W(\Lambda)) \\
 P(w) &\equiv g\phi_0(w)w + \frac{1}{2}m_{\phi}^2\phi_0(w)^2
             +\frac{\lambda}{4}\phi_0(w)^4,
\end{aligned}
\end{equation}
using (\ref{effective-hamiltonian}). When computing $K(\Lambda)$ and $W(\Lambda)$ above, we dropped the terms that diverge without cutoff, following Assumption 1. The equalities $K(0)=W(0)=0$ derived from Assumption 1 in turn ensure that $\rho(\Lambda=0)=0$. Thus, Assumption 1 can be rephrased as the postulate that the vacuum energy in the absence of fermions exactly vanishes.

Next, we calculate the first order term in $\hat{H}$. Let $\Lambda_0$ be the value of $\Lambda$ such that $\rho(\Lambda_0)$ is a local minimum of $\rho(\Lambda)$ if it exists. For $\Lambda=\Lambda_0$, the expansion of the Hamiltonian by fermion bilinears up to the first order has a simple form,
\begin{equation}
\label{hamiltonian-expansion}
\begin{aligned}
  \hat{H} &\simeq \rho(\Lambda_0)V
          +\int d^3x\left[\frac{1}{2}\bar{\psi}\left(-i\bfgamma\cdot\bfnabla+m\right)\psi
                 -K(\Lambda_0)\right] \\
          &\quad +g\phi_0(W(\Lambda_0))
                  \int d^3x\left(\frac{\bar{\psi}\psi}{2}-W(\Lambda_0)\right) \\
          &= \rho(\Lambda_0)V
             +\sum_{|\bfp|\leq\Lambda_0}E_{|\bfp|}a_{\bfp}\cdot a_{\bfp}^{\dagger}
             +\sum_{|\bfp|>\Lambda_0}E_{|\bfp|}a_{\bfp}^{\dagger}\cdot a_{\bfp} \\
          &\quad -g\sum_{\bfp}\frac{\phi_0(W(\Lambda_0))|\bfp|}{2\omega_{|\bfp|}}
                     \left(a^{\dagger}_{\bfp}\cdot Ma^{\dagger}_{-\bfp}
                           +a_{-\bfp}\cdot M^{\dagger}a_{\bfp}\right),
\end{aligned}
\end{equation}
where we defined $M\equiv (p_i\cdot\sigma_i)(i\sigma_2)/|\bfp|$ using the Pauli matrices $\sigma_i$ and
\begin{equation}
\label{energy-cost}
  E_p \equiv \left|\omega_p+g\phi_0(W(\Lambda_0))\frac{m}{\omega_p}\right|
      = \omega_p\left|1-\left(\frac{\omega_{\Lambda_0}}{\omega_p}\right)^2\right|.
\end{equation}
In (\ref{energy-cost}), we used the extremality condition
\begin{equation}
\label{minimum-condition}
 \rho'(\Lambda_0) = \frac{\Lambda_0^2}{\pi^2}
       \left[\omega_{\Lambda_0}+g\phi_0(W(\Lambda_0))\frac{m}{\omega_{\Lambda_0}}\right]=0.
\end{equation}

Finally, we decompose the first order part of (\ref{hamiltonian-expansion}) into $\hat{H}_0$ and $\hat{H}_I$. We wish to define $\hat{H}_0$ and $\hat{H}_I$ so that they give a free Hamiltonian and interaction part of $\hat{H}$, respectively. Such a definition is naturally given by
\begin{equation}
\label{exchange-interaction}
\begin{aligned}
  \hat{H}_0 &\equiv \sum_{|\bfp|\leq\Lambda_0}E_{|\bfp|}a_{\bfp}\cdot a_{\bfp}^{\dagger}
                   +\sum_{|\bfp|>\Lambda_0}E_{|\bfp|}a_{\bfp}^{\dagger}\cdot a_{\bfp}, \\
  \hat{H}_I
    &\equiv \frac{\omega_{\Lambda_0}^2}{m}
            \sum_{\bfp}\frac{|\bfp|}{2\omega_{|\bfp|}}
                       \left(a^{\dagger}_{\bfp}\cdot Ma^{\dagger}_{-\bfp}
                             +a_{-\bfp}\cdot M^{\dagger}a_{\bfp}\right).
\end{aligned}
\end{equation}
Since all particle states with $|\bfp|\leq\Lambda_0$ are occupied in $|\Lambda_0\rangle$, excited states are obtained by applying $a_{\bfp}$ with $|\bfp|\leq\Lambda_0$ or $a_{\bfp}^{\dagger}$ with $|\bfp|>\Lambda_0$ on $|\Lambda_0\rangle$. We refer to these excitations as ``excited particles'' if $|\bfp|>\Lambda_0$, or ``holes'' if $|\bfp|\leq\Lambda_0$. As (\ref{exchange-interaction}) indicates, $E_p\geq 0$ describes the dispersion relation of the excited particles and holes. The interaction part $\hat{H}_I$ contains pair-creation and pair-annihilation interactions of fermions.

\subsection{Thermal history of the DE sector}

In this subsection, we study the behavior of the energy density of the DE sector from the era of reheating to the present Universe. Let us decompose $\rho_T$ and $p_T$ into the two components to which the SM and DE sectors contribute,
\begin{equation}
 \label{energy-decomposition}
 \begin{aligned}
  \rho_T &= \rho_{SM}+\rho, \\
  p_T &= p_{SM}+p. \\
 \end{aligned}
\end{equation}
The notations used in the r.h.s. of (\ref{energy-decomposition}) will be self-evident.

In the three subsequent sections, we show the following results in case of $T\ll 10^{-3}~\textrm{eV}$.
\begin{description}
 \item[Section \ref{section-metastable}] The function $\rho(\Lambda)$ has a positive local minimum $\rho(\Lambda_0)\sim (10^{-3}~\textrm{eV})^4$ at $\Lambda=\Lambda_0\sim 10^{-3}~\textrm{eV}$ for a certain parameter range.
 \item[Section \ref{section-pair-production}] The equation of state for the DE sector is given by $p\simeq -\rho$.
 \item[Section \ref{section-fate}] The decay rate of the metastable state is so small that the metastable state is expected not to decay into the ground state for a period longer than the age of the Universe. 
\end{description}
Let us accept these results in advance for the moment, and derive the dependence of $\rho$ on $T$. The result of section \ref{section-metastable} suggests that there is an equilibrium at $\Lambda=\Lambda_0$. We refer to $n_0\equiv2\int_{|\bfp|\leq\Lambda_0} d^3p(2\pi)^{-3}=\Lambda_0^3/\pi^2$ as the ``equilibrium number density'' of fermions. If $T\gg 10^{-3}~\textrm{eV}$, both $\psi$ and $\phi$ behave as relativistic particles. Then, the number density $n$ of fermions is given by $n=(3\zeta(3)/4\pi^2)g_F T^3$, where $g_F=2$ for Majorana fermion case \cite{kolb-turner}. As the Universe cools down and $T$ approaches to $10^{-3}~\textrm{eV}$, $n$ gets close to the equilibrium number density $n_0=\Lambda_0^3/\pi^2$. When $n\simeq n_0$, the DE sector will be trapped at the local minimum $\rho(\Lambda_0)$. Even if the Universe further evolves, the DE sector continues staying at $\rho(\Lambda_0)$. In summary, the energy density $\rho$ of the DE sector depends on $T$ as follows:
\begin{equation}
 \rho = \left\{
        \begin{array}{ll}
	 (\pi^2/30)(7g_F/8+1)T^4 & (T\gg 10^{-3}~\textrm{eV}) \\
         \rho(\Lambda_0) & (T\ll 10^{-3}~\textrm{eV})
	\end{array}
        \right..
\end{equation}

Since the SM sector and the DE sector are almost decoupled, the entropy of the SM sector is not transferred to the DE sector as $T_{SM}$ decreases. Therefore, when $T\gtrsim 10^{-3}~\textrm{eV}$, radiation or matter in the SM sector dominates $\rho_T$. In particular, it is worth stressing that $\rho$ only accounts for a tiny fraction of $\rho_T$, not spoiling the success of the standard cosmology including the Big Bang nucleosynthesis. On the other hand, the DE sector begins dominating $\rho_T$ when $\rho_{SM}\simeq \rho(\Lambda_0)$ at some temperature $T <10^{-3}~\textrm{eV}$. Since $\rho_T\simeq \rho=\rho(\Lambda_0)$ and $p_T\simeq p=-\rho(\Lambda_0)$ for $\rho_{SM}\ll \rho(\Lambda_0)$, an accelerating expansion of the Universe derived from (\ref{friedmann}),
\begin{equation}
 a \propto \exp(H_0 t),\quad H_0 \equiv \sqrt{\frac{\rho(\Lambda_0)}{3M_P^2}},
\end{equation}
begins recently at the era $\rho_{SM}\simeq \rho(\Lambda_0)\sim (10^{-3}~\textrm{eV})^4$.

\section{metastable state and dark energy}
\label{section-metastable}

\begin{figure}
 \includegraphics[scale=0.07]{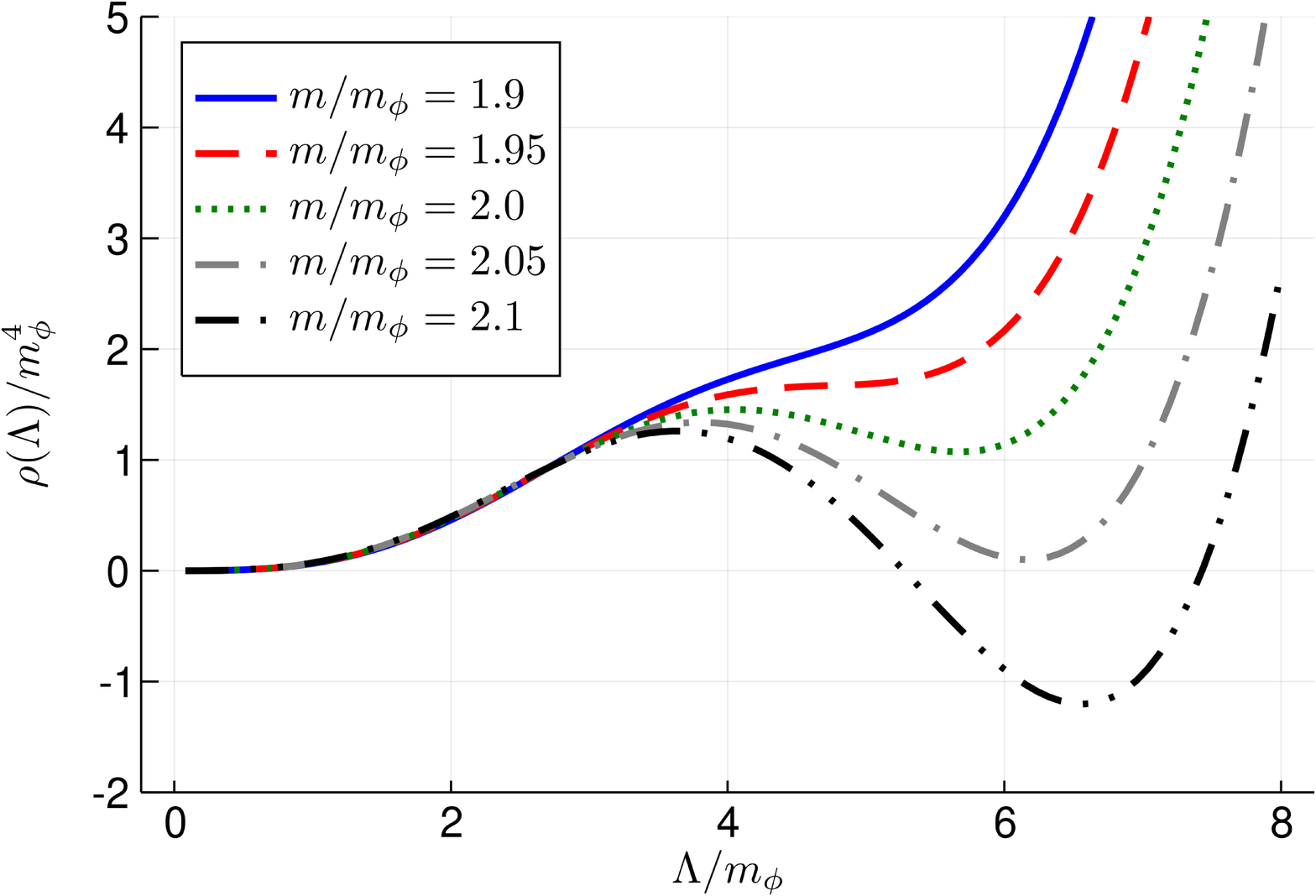}
 \caption{\label{energy-density1}Functional dependence of $\rho(\Lambda)$ on $\Lambda\in[0,8m_{\phi}]$ for $g=3$ and $\lambda=0.01$. The horizontal and vertical axis stands for $\Lambda/m_{\phi}$ and $\rho(\Lambda)/m_{\phi}^4$, respectively. The graphs for five representative values of $m/m_{\phi}\in[1.9,2.1]$ are shown. A positive local minimum exists at least for $2.0\leq m/m_{\phi}\leq 2.05$.}
\end{figure}

Let us investigate if there exists a positive local minimum of $\rho(\Lambda)$, which is a potential candidate for the dark energy. All possible parameter space will not be exhaustively searched, but we just attempt to find a parameter region which is consistent with the observation of the vacuum energy density. We consider only the case $g=3$ and $\lambda=0.01$ as an illustrative example of the model. As depicted in FIG. \ref{energy-density1}, a local minimum exists for a range of model parameters. If we demand $\rho(\Lambda_0)=(10^{-3}~\textrm{eV})^4$, the fermion mass takes values $m=(2.0\times 10^{-3}\mathchar`- 3.7\times 10^{-3})~\textrm{eV}$ for $m/m_{\phi}=2.0\mathchar`-2.05$. In principle, there exists no upper bound for $m$, since $\rho(\Lambda_0)/m_{\phi}^4$ can be made arbitrarily small if we appropriately adjust $m/m_{\phi}$, as FIG. \ref{energy-density1} indicates. However, more fine-tuning of $m/m_{\phi}$ is necessary to obtain larger values of $m$ consistent with $\rho(\Lambda_0)=(10^{-3}~\textrm{eV})^4$. From now on, the sphere in the momentum space defined by $|\bfp|=\Lambda_0$ is referred to as the Fermi surface.

A physical explanation of the reason why $\rho(\Lambda)$ has a local minimum is as follows. If $\Lambda$ is sufficiently small, $\rho(\Lambda)$ is approximated by
\begin{equation}
\label{rho-approximation}
  \rho(\Lambda) = K(\Lambda) - \frac{g^2}{2m_{\phi}^2}W(\Lambda)^2,
\end{equation}
as derived from (\ref{approximate-solution}) and (\ref{internal-energy-density}). The second term of the r.h.s. of (\ref{rho-approximation}) is an attractive potential between fermions induced by one scalar exchange. As $\Lambda$ increases and the decrease in the negative attractive potential dominates the increase in $K(\Lambda)$, $\rho(\Lambda)$ begins decreasing at its local maximum shown in FIG. \ref{energy-density1}. However, as $\Lambda$ increases further, the higher order terms in $P(W(\Lambda))$ contributes to $\rho(\Lambda)$ by a positive amount. This positive contribution by $P(W(\Lambda))$ is regarded as a repulsive interaction between fermions. Because of this repulsive potential, $\rho(\Lambda)$ begins increasing again. In summary, at the local minimum $\Lambda=\Lambda_0$, the increase in the kinetic energy plus repulsive potential exactly balances the decrease in the attractive potential, when a small number of fermions are added on the Fermi surface.

So far, we have only found that there exists a parameter range in which the state $|\Lambda_0\rangle$ is metastable along the subspace ${\cal H}'$ of ${\cal H}$. However, it is also possible to show that the state $|\Lambda_0\rangle$ is metastable along the directions in ${\cal H}$ not parallel to ${\cal H}'$. The remainder of this section is devoted to showing this statement.

Let us consider an excited particle state $|\bfp,\sigma\rangle\equiv a^{\dagger}(\bfp,\sigma)|\Lambda_0\rangle$ with $|\bfp|>\Lambda_0$, but not $|\bfp|\gg\Lambda_0$. This state has an excitation energy $E_{|\bfp|}$, as can be seen from (\ref{exchange-interaction}). Note that the excitation energy $E_{|\bfp|}$ vanishes at the Fermi surface $|\bfp|=\Lambda_0$. Therefore, the excited particle exhibits tachyon-like behavior and decays into many fermions with nearly vanishing energy as follows. Because $E_{\Lambda_0}=0$, there must exist a state
\begin{equation}
\label{excited-state}
 |n,\Lambda_0\rangle\equiv
  \prod_{|\bfp|\leq\Lambda_0,\sigma}a(\bfp,\sigma)^{n_{{\bfp},\sigma}}
  \prod_{|\bfp|>\Lambda_0,\sigma}a^{\dagger}(\bfp,\sigma)^{n_{{\bfp},\sigma}}
  |\Lambda_0\rangle
\end{equation}
which has the same four-momentum and spin as those of $|\bfp,\sigma\rangle$, and satisfies $n_{{\bfp},\sigma}=1$ only if $\Lambda_0-\delta\Lambda_0<|\bfp|<\Lambda_0+\delta\Lambda_0$ with a small $\delta\Lambda_0$. Recall that the Yukawa coupling of the fermion with high energy modes of the scalar has been ignored by integrating out the time-dependent component of $\phi$ in (\ref{lag}). Through such a coupling, the state $|\bfp,\sigma\rangle$ immediately transitions to the multi-particle state $|n,\Lambda_0\rangle$. In a similar way, the state $a(\bfp,\sigma)|\Lambda_0\rangle$ corresponding to a hole with $|\bfp|\leq\Lambda_0$ instantaneously decays into a state of the form $|n,\Lambda_0\rangle$. Since the state $|\Lambda_0\rangle$ has a macroscopically large number of fermions, it is almost impossible to distinguish $|n,\Lambda_0\rangle$ from $|\Lambda_0\rangle$. Thus, we conclude that excited particles and holes are unstable and eventually transition to a state which is virtually identical with $|\Lambda_0\rangle$.

\section{eternal pair-production of fermions in the expanding universe}
\label{section-pair-production}

It is straightforward to show that the state $|\Lambda_0\rangle$ satisfies the equation of state $w\equiv p/\rho\simeq -1$, in consistent with the latest observation \cite{planck2018-6}. The internal energy $U$ is written as $U=TS-pV+\mu N$, where $\mu$ is the chemical potential and $N$ is the number of fermions. Since $N$ is not conserved, $\mu=0$ at the equilibrium corresponding to the metastable state $|\Lambda_0\rangle$. Hence $\rho+p=(U+pV)/V\simeq 0$ follows from Assumption 2. As a result, the energy density of the state $|\Lambda_0\rangle$ depends on the scale factor $a$ of the Universe as $\rho(\Lambda_0)\propto a^{-3(w+1)}=a^0$. However, it might sound strange that the energy density of particles does not vary as the Universe expands. In fact, the energy density of free relativistic (resp. non-relativistic) particles is diluted as $\rho\propto a^{-4}$ (resp. $\rho\propto a^{-3}$) as $a$ increases.

To understand why $\rho(\Lambda_0)$ remains constant, suppose that the scale factor $a$ increases by a small amount $\delta a$. Then, the state $|\Lambda_0\rangle$ is red-shifted to $|\Lambda_0-\Delta\Lambda\rangle$, with $\Delta\Lambda=(\delta a/a)\Lambda_0$. Now we can show that the state $|\Lambda_0-\Delta\Lambda\rangle$ is unstable and transitions to the state $|\Lambda_0+\Delta\Lambda\rangle$, as follows. The excitation energy of the state $a^{\dagger}(\bfp,\sigma)|\Lambda_0-\Delta\Lambda\rangle$ relative to $|\Lambda_0-\Delta\Lambda\rangle$ can be read off from the commutation relation
\begin{equation}
  [\hat{H}_0, a^{\dagger}(\bfp,\sigma)] =
  \left\{
         \begin{array}{ll}
	  -E_{|\bfp|}a^{\dagger}(\bfp,\sigma) & (|\bfp|\leq\Lambda_0) \\
          E_{|\bfp|}a^{\dagger}(\bfp,\sigma) & (|\bfp|>\Lambda_0)
	 \end{array}
  \right..
\end{equation}
Note that $E_p\simeq(2\Lambda_0/\omega_{\Lambda_0})|p-\Lambda_0|$ for $p\simeq\Lambda_0$, from (\ref{energy-cost}). Therefore, the fermions with $\Lambda_0-\Delta\Lambda<|\bfp|<\Lambda_0$ have negative energy $-E_{|\bfp|}$, which is canceled by the positive energy $E_{2\Lambda_0-|\bfp|}=E_{|\bfp|}$ of the fermions with momentum $2\Lambda_0-|\bfp|\in[\Lambda_0, \Lambda_0+\Delta\Lambda]$. This implies that the state $|\Lambda_0-\Delta\Lambda\rangle$ immediately decays into $|\Lambda_0+\Delta\Lambda\rangle$ through the pair-production of fermions with momentum $\bfp$ in the interval $[\Lambda_0-\Delta\Lambda,\Lambda_0+\Delta\Lambda]$. The fact that the transition $|\Lambda_0-\Delta\Lambda\rangle\rightarrow|\Lambda_0+\Delta\Lambda\rangle$ is possible is attributed to the equality $E_{|\bfp|}\simeq 0$ in the vicinity of the Fermi surface, which follows from $\rho'(\Lambda_0)=0$. The negative term $g\phi_0(W(\Lambda_0))(m/\omega_{\Lambda_0})$ in (\ref{minimum-condition}) is the potential energy felt by one fermion in the background of other fermions. Therefore, the condition $\rho'(\Lambda_0)=0$ given by (\ref{minimum-condition}) indicates that the kinetic energy $\omega_{\Lambda_0}$ of an excited particle on the Fermi surface is canceled by the negative potential energy generated by the particles in $|\Lambda_0\rangle$.

\begin{figure}
 \includegraphics[scale=0.5]{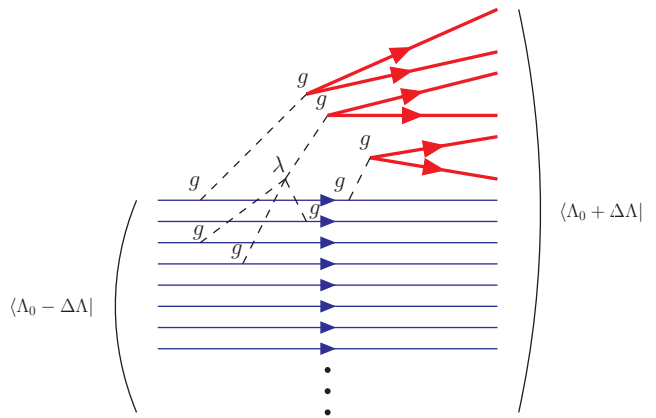}
 \caption{\label{feynman}The transition process from $|\Lambda_0-\Delta\Lambda\rangle$ to $|\Lambda_0+\Delta\Lambda\rangle$. Straight and dotted lines stand for the fermion $\psi$ and scalar propagator $\Delta(\bfx-\bfy)$, respectively. This process is kinematically allowed, since the kinetic energy of the pair-produced fermions (thick straight lines) is exactly canceled by the negative potential energy generated by the fermions in $|\Lambda_0-\Delta\Lambda\rangle$.}
\end{figure}

We depict a Feynman diagram for the transition process $|\Lambda_0-\Delta\Lambda\rangle\rightarrow|\Lambda_0+\Delta\Lambda\rangle$ in FIG. \ref{feynman}. The pair-production of fermions $\psi$ occurs via the interaction $\hat{H}_I$. Therefore, after the states with momentum $|\bfp|\in [\Lambda_0-\Delta\Lambda,\Lambda_0]$ become unoccupied due to red-shift, other fermions instantaneously fill up the states with momentum $|\bfp|\in[\Lambda_0-\Delta\Lambda,\Lambda_0+\Delta\Lambda]$. Then, $\Lambda$ is red-shifted again from $\Lambda_0+\Delta\Lambda$ to $\Lambda_0-\Delta\Lambda$. The process of red-shift followed by fermion pair-production never terminates as long as the Universe expands. As a consequence, $|\Lambda\rangle$ fluctuates around the Fermi surface with a small amplitude $\Delta\Lambda$, as if no red-shift occurs. The fluctuation amplitude $\Delta\Lambda$ must be evaluated by a careful analysis involving the Hubble parameter $H\equiv\dot{a}/a$ and the transition amplitude of $|\Lambda_0+\Delta\Lambda\rangle\rightarrow|\Lambda_0-\Delta\Lambda\rangle$. However, we may conclude that $\Delta\Lambda$ is sufficiently small even without such an analysis, since the expansion rate of the present Universe, $H_0\sim 10^{-33}~\textrm{eV}$, is extremely smaller than the characteristic scale of the model $\sim 10^{-3}~\textrm{eV}$.

\section{fate of metastable dark energy}
\label{section-fate}

In this section, we evaluate the probability of the metastable state $|\Lambda_0\rangle$ penetrating the potential barrier and migrating to the true vacuum $|0\rangle$. The aim of this evaluation is to confirm that the probability is so small that the present Universe is highly likely to be in the metastable state with the dark energy. The decay rate of a false vacuum in a scalar field theory has been calculated by Coleman and Callan in \cite{coleman}. We first briefly review the methodology presented in \cite{coleman}, and then apply that  methodology to our model.

The authors of \cite{coleman} considered the Euclidean action for a real scalar field $\varphi$,
\begin{equation}
 S_{\varphi} = \int d^4x
      \left[\frac{1}{2}\left(\frac{\partial\varphi}{\partial\tau}\right)^2
            +\frac{1}{2}|\bfnabla\varphi|^2+U(\varphi)\right],
\end{equation}
where $U(\varphi)$ is a potential with the global minimum at $\varphi=\varphi_-$ and a local minimum $\varphi=\varphi_+$ separated by a potential barrier. It is assumed that the local minimum is zero, namely $U(\varphi_+)=0$. The global and local minima correspond to the true and a false vacuum of the system, respectively. Let us define a radial coordinate $r\equiv(\tau^2+|\bfx|^2)^{1/2}$. The equation of motion derived from the action $S_{\varphi}$ admits a $O(4)$ symmetric solution $\bar{\varphi}(r)$ with the boundary condition
\begin{equation}
 \lim_{r\rightarrow\infty}\bar{\varphi}(r) = \varphi_+, \qquad
 \left.\frac{\partial\bar{\varphi}(r)}{\partial r}\right|_{r=0} = 0.
\end{equation}
In addition to the trivial solution $\bar{\varphi}(r)=\varphi_+$, there is a non-trivial solution corresponding to the vacuum transition from $\varphi_+$ to $\varphi_-$. Such a solution is referred to as the ``bounce solution''. We define ``the bounce action'' $B$ by the value of the action $S_{\varphi}$ under the bounce solution $\varphi=\bar{\varphi}$, as
\begin{equation}
 B \equiv 2\pi^2\int_0^{\infty}r^3dr
          \left[\frac{1}{2}\left(\frac{\partial\bar{\varphi}}{\partial r}\right)^2
                +U(\bar{\varphi})\right].
\end{equation}
The decay rate of the false vacuum per unit volume is then given by
\begin{equation}
 \label{tunneling-prob}
 \Gamma/V \sim A\exp(-B),
\end{equation}
where $A$ is a quantity with dimension $[\textrm{mass}]^4$ that depends on the detail of the model.

For a specific form of the potential $U(\varphi)$, an analytic expression for $B$ can be obtained. An example of such potential is
\begin{equation}
 \label{quartic-potential}
 U(\varphi) = \frac{\kappa}{4}\varphi^2(\varphi+2b)^2-\frac{\epsilon}{2b}(\varphi+2b),
\end{equation}
where $\kappa$, $b$, and $\epsilon$ are positive constants. If $\epsilon$ is sufficiently small, $\varphi_-\simeq 0$ and $\varphi_+\simeq -2b$. An approximate expression for the bounce action under small $\epsilon$ has been calculated in \cite{coleman} as
\begin{equation}
 \label{bounce-action}
 B\simeq\frac{2\pi^2\kappa^2b^{12}}{3\epsilon^3}.
\end{equation}

Next, let us utilize the above methodology to estimate the tunneling probability of our model. To this end, we have to return to the Lagrangian $L$ in (\ref{lag}) and integrate out the fermion $\psi$, to obtain a model that contains only the scalar $\phi$. Since $L$ is bilinear in terms of $\psi$, it is possible to exactly carry out the functional integration with respect to $\psi$. However, such a naive integration is inappropriate for our current purpose, as we will see now. The effective action for $\phi$ obtained by the path integral $\int[{\cal D}\bar{\psi}{\cal D}\psi]\exp(i\int dtL)$ contains the term
\begin{equation}
 \label{one-loop}
 \ln\textrm{Det}\left[\frac{1}{2}(i\fslash{\partial}-m)-\frac{g}{2}\phi\right].
\end{equation}
The expansion of (\ref{one-loop}) in terms of $\phi$ gives the self-interaction terms of $\phi$ mediated by the virtual fermion circulating around an internal loop. In other words, fermions only appear as unobservable virtual particles in the system under consideration. However, as revealed in the preceding sections, the Universe in the metastable state is filled with observable \textit{on-shell} fermions, not with intermediate off-shell fermions. Therefore, the exact functional integration does not properly describes the dark energy predicted by our model.

To find a pure scalar field theory suitable for our goal, let us write down the Hamiltonian density $\hat{{\cal H}}$ derived directly from $L$ as
\begin{equation}
\begin{aligned}
 \hat{{\cal H}} &\equiv \frac{1}{2}\pi_{\phi}^2+\frac{1}{2}|\bfnabla\phi|^2 + V(\phi,\psi), \\
 V(\phi,\psi) &\equiv \frac{1}{2}m_{\phi}^2\phi^2+\frac{\lambda}{4}\phi^4 \\
   &\quad +\frac{1}{2}\bar{\psi}\left(-i\bfgamma\cdot\bfnabla+m\right)\psi
                                +\frac{g}{2}\phi\bar{\psi}\psi,
\end{aligned}
\end{equation}
where $\pi_{\phi}$ is the conjugate momentum of $\phi$. To appropriately eliminate $\psi$ from $\hat{{\cal H}}$, it is helpful to use a mean field approximation again. In the present case, we take the expectation value of $V(\phi,\psi)$ with the state $|\Lambda\rangle$ and minimize it with respect to $\Lambda$. The effective potential for $\phi$ defined this way is given by
\begin{equation}
\label{effective-potential}
\begin{aligned}
  V_{\textrm{eff}}(\phi) &\equiv \min_{\Lambda}\langle\Lambda|V(\phi,\psi)|\Lambda\rangle \\
 &= \min_{\Lambda}\left[K(\Lambda)+g\phi W(\Lambda)\right]
                      +\frac{1}{2}m_{\phi}^2\phi^2+\frac{\lambda}{4}\phi^4.
\end{aligned}
\end{equation}
The minimum condition in (\ref{effective-potential}) reads
\begin{equation}
\label{minimum-condition2}
 \Lambda^2\left(\omega_{\Lambda}+g\phi\frac{m}{\omega_{\Lambda}}\right)=0.
\end{equation}
The solution to (\ref{minimum-condition2}) that gives the minimum of (\ref{effective-potential}) is given by $\Lambda=\Lambda_{\phi}\equiv\sqrt{m|g\phi+m|}$ for $\phi<-m/g$ and $\Lambda=0$ for $\phi\geq -m/g$. As a consequence, we obtain
\begin{equation}
 V_{\textrm{eff}}(\phi)
  =\left\{
   \begin{array}{ll}
    \begin{array}{l}
     K(\Lambda_{\phi})+g\phi W(\Lambda_{\phi}) \\
     +\frac{1}{2}m_{\phi}^2\phi^2+\frac{\lambda}{4}\phi^4
    \end{array}
     & \left(\phi<-m/g\right) \\
     & \\
    \frac{1}{2}m_{\phi}^2\phi^2+\frac{\lambda}{4}\phi^4 & \left(\phi\geq -m/g\right)
   \end{array}
   \right..
\end{equation}

\begin{figure}
 \includegraphics[scale=0.07]{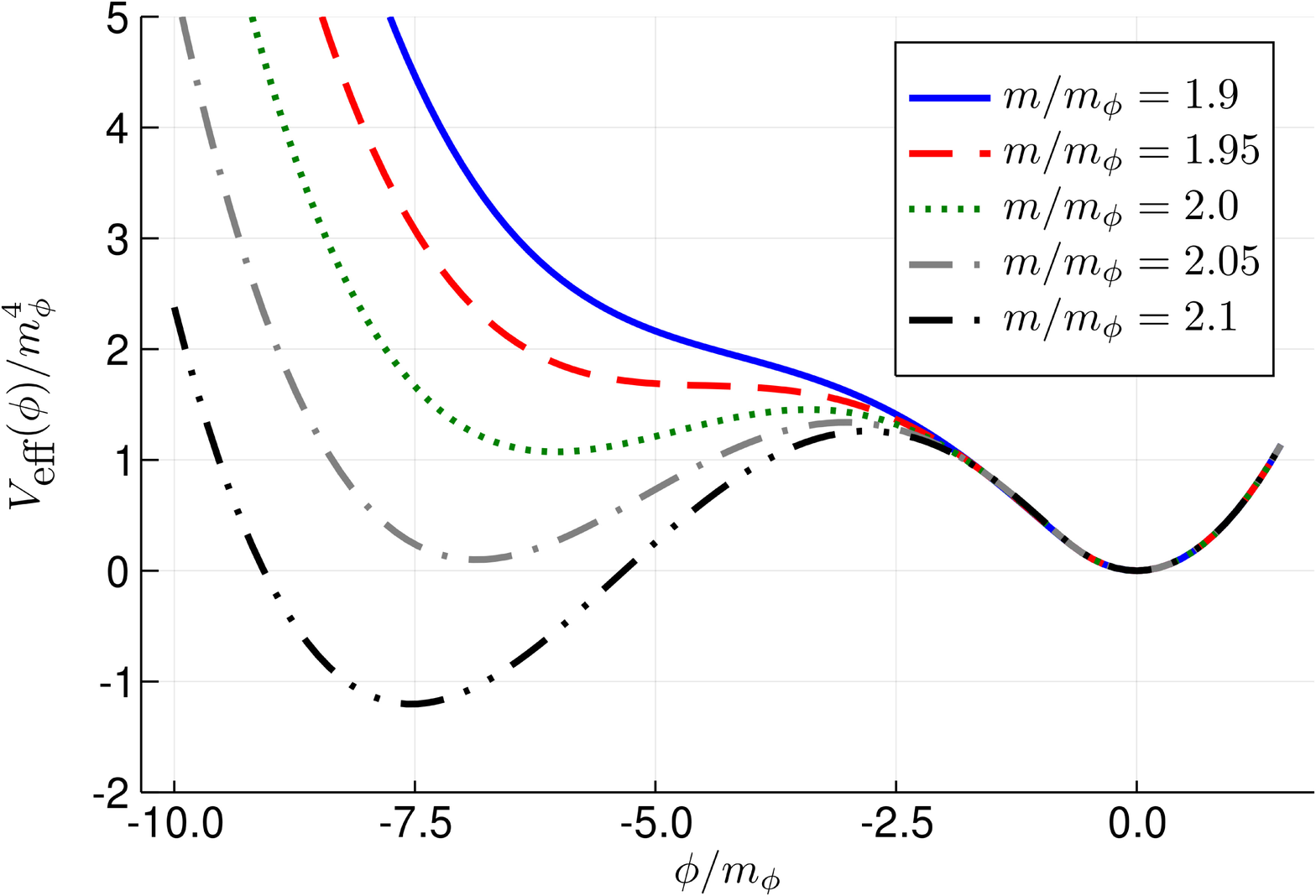}
 \caption{\label{fig-effective-potential}The scalar effective potential $V_{\textrm{eff}}(\phi)$ for the same parameter range as that given in FIG. \ref{energy-density1}. At least for $2.0\leq m/m_{\phi}\leq 2.05$, there is a positive local minimum corresponding to the dark energy, which is separated from the true vacuum at $\phi=0$ by a potential barrier.}
\end{figure}

We depict the graphs of $V_{\textrm{eff}}(\phi)$ in FIG. \ref{fig-effective-potential} for the same parameter range as the one used for calculating $\rho(\Lambda)$ in FIG. \ref{energy-density1}. The functional dependence of $V_{\textrm{eff}}(\phi)$ on $\phi$ resembles that of $\rho(\Lambda)$, as all the global or local extrema of $V_{\textrm{eff}}(\phi)$ have the same value as those of $\rho(\Lambda)$. This is not accidental. In fact, $\rho(\Lambda)$ and $V_{\textrm{eff}}(\phi)$ are obtained by minimizing the same function $\langle\Lambda|V(\phi,\psi)|\Lambda\rangle$ with $\phi$ and $\Lambda$, respectively (recall the expression of $\rho(\Lambda)$ in (\ref{internal-energy-density})). The local minimum of $V_{\textrm{eff}}(\phi)$ with $\phi<0$ and $m/m_{\phi}\lesssim 2$ arises from the negative attractive potential between fermions. This implies that fermions are appropriately treated as observable particles (not virtual particles), in the definition (\ref{effective-potential}) of $V_{\textrm{eff}}(\phi)$.

Finally, let us calculate the tunneling probability in the case $m/m_{\phi}=2.05$. Note that, in that case, $V_{\textrm{eff}}(\phi)$ can be well approximated by a quartic polynomial $U(\varphi)$ in (\ref{quartic-potential}) with small $\epsilon$. This rather bold approximation is admissible since our aim is just to obtain a rough estimate of the tunneling probability
\footnote{
Actually, $V_{\textrm{eff}}(\phi)$ differs from $U(\phi)$ by a constant $\epsilon$. This difference does not affect the evaluation of the tunneling probability.
}
. Let $\phi_m$ and $\phi_+<0$ be the value of $\phi$ that give the local maximum and minimum of $V_{\textrm{eff}}(\phi)$, respectively. We determine $\kappa$, $b$, and $\epsilon$ by assuming that the following three values are the same between $U(\varphi)$ and $V_{\textrm{eff}}(\phi)$: 1) local maximum, 2) horizontal position of the local minimum, and 3) difference between the global minimum and the local minimum. Then, we obtain three equalities as follows:
\begin{equation}
\label{correspondence}
 \kappa b^4/4 \simeq V_{\textrm{eff}}(\phi_m),\quad -2b \simeq \phi_+, \quad
 \epsilon \simeq V_{\textrm{eff}}(\phi_+).
\end{equation}
For $m/m_{\phi}=2.05$, we have $\phi_m\simeq -3m_{\phi}$ and $\phi_+\simeq -7m_{\phi}$. From a dimensional analysis, we make another bold assumption about the coefficient $A$ as
\begin{equation}
\label{coefficent-A}
 A=C{\cal M}^4,
\end{equation}
where $C$ is a constant not significantly different from unity and ${\cal M}=10^{-3}~\textrm{eV}$ is the characteristic scale of the model. The tunneling probability we wish to compute is obtained by multiplying on $\Gamma/V$ the age $H_0^{-1}$ and the volume $H_0^{-3}$ of the observable Universe. Hence, from (\ref{tunneling-prob}), (\ref{bounce-action}), (\ref{correspondence}), and (\ref{coefficent-A}), the tunneling probability is roughly evaluated as
\begin{equation}
\label{tunneling-prob2}
 \frac{A}{H_0^4}\exp(-B) \sim C\times 10^{-10^7}.
\end{equation}
The extremely small probability (\ref{tunneling-prob2}) indicates that the dark energy is stable against vacuum decay at least for a period comparable to the present age of the Universe.

\section{dirac fermion case}
\label{section-dirac}

In this section, we consider the case where fermions are described by Dirac spinor $\psi$. Our goal is to show that there is a metastable state for a certain parameter set. Since the Dirac fermion case can be treated in a way similar to the Majorana case, we try to keep our discussion in this section as brief as possible.

The Lagrangian is given by
\begin{align}
 \label{lag-dirac}
\begin{aligned}
 L_D &\equiv \int d^3x\left[
              \bar{\psi}\left(i\fslash{\partial}-m\right)\psi-g\phi\bar{\psi}\psi\right. \\
   & \qquad\qquad\quad +\left.
              \frac{1}{2}\partial_{\mu}\phi\partial^{\mu}\phi-\frac{1}{2}m_{\phi}^2\phi^2
              -\frac{\lambda}{4}\phi^4\right],
\end{aligned}
\end{align}
where we used the same symbols for four model parameters as those in the Majorana case. By using the functional $\phi_0$ defined in (\ref{solution}), we write down the corresponding low energy effective Hamiltonian as
\begin{equation}
 \label{effective-hamiltonian-dirac}
\begin{aligned}
 \hat{H}_D &\equiv \int d^3x\left[\bar{\psi}\left(-i\bfgamma\cdot\bfnabla+m\right)\psi
                                +g\phi_0(\bar{\psi}\psi)\bar{\psi}\psi\right. \\
   &\qquad\qquad\quad
           +\frac{1}{2}\phi_0(\bar{\psi}\psi)(-\bfnabla^2+m_{\phi}^2)\phi_0(\bar{\psi}\psi) \\
   &\qquad\qquad\quad
            \left.+\frac{\lambda}{4}\phi_0(\bar{\psi}\psi)^4\right].
\end{aligned}
\end{equation}
The Dirac field $\psi$ describes particles and anti-particles. Denoting the creation operators of particles and anti-particles by $b^{\dagger}_{\bfp,\sigma}$ and $d^{\dagger}_{\bfp,\sigma}$ respectively
\footnote{
Similarly to the footnote \ref{vector-notation}, we use a simple notation for creation and annihilation operators such as $b_{\bfp}^{\dagger}=(b^{\dagger}_{\bfp,+},b^{\dagger}_{\bfp,-})^T$ below.
}, we can expand $\psi$ as
\begin{equation}
\label{psi_dirac}
\begin{aligned}
  \psi(\bfx) &= \sum_{\bfp,\sigma}\frac{1}{\sqrt{2\omega_{|\bfp|}V}}
                \left[u(\bfp,\sigma)e^{i\bfp\cdot\bfx}b_{\bfp,\sigma}\right. \\
             & \qquad\qquad\qquad\qquad
                      +\left.v(\bfp,\sigma)e^{-i\bfp\cdot\bfx}d^{\dagger}_{\bfp,\sigma}\right].
\end{aligned}
\end{equation}
Similarly to section \ref{section-model}, instead of the entire Hilbert space of the model, we focus on a subspace ${\cal H}'_D$ of it defined by
\begin{equation}
 \label{hilbert_dirac}
 {\cal H}'_D \equiv
  \left\{\left.|\Lambda\rangle
                =\left[\prod_{|\bfp|\leq\Lambda,\sigma=\pm}b^{\dagger}_{\bfp,\sigma}d^{\dagger}_{\bfp,\sigma}\right]
                 |0\rangle
         \right|\Lambda\in\mathbf{R}_{\geq 0}\right\},
\end{equation}
where the vacuum $|0\rangle$ satisfies $b_{\bfp,\sigma}|0\rangle=d_{\bfp,\sigma}|0\rangle=0$. The form of $|\Lambda\rangle$ in ${\cal H}'_D$ indicates that particles and anti-particles are degenerate in the Fermi surface with a common radius $\Lambda$. We also define the expectation values $K_D(\Lambda)$ and $W_D(\Lambda)$ as
\begin{equation}
 \label{kinetic-dirac}
\begin{aligned}
  K_D(\Lambda)
  &\equiv \langle\Lambda|\bar{\psi}\left(-i\bfgamma\cdot\bfnabla+m\right)\psi
          |\Lambda\rangle = 2K(\Lambda), \\
  W_D(\Lambda) &\equiv \langle\Lambda|\bar{\psi}\psi|\Lambda\rangle = 2W(\Lambda).
\end{aligned}
\end{equation}
Under a mean field approximation, the Hamiltonian $\hat{H}_D$ is approximated in the vicinity of the state $|\Lambda_0\rangle$ as
\begin{equation}
\label{exchange-interaction-dirac}
\begin{aligned}
  \hat{H}_D &\simeq \rho_D(\Lambda_0)V + \hat{H}_{D0} + \hat{H}_{DI}, \\
  \rho_D(\Lambda) &\equiv K_D(\Lambda) + P(W_D(\Lambda)), \\
  \hat{H}_{D0} &\equiv \sum_{|\bfp|\leq\Lambda_0}E_{|\bfp|}b_{\bfp}\cdot b_{\bfp}^{\dagger}
                   +\sum_{|\bfp|>\Lambda_0}E_{|\bfp|}b_{\bfp}^{\dagger}\cdot b_{\bfp} \\
            &\quad +\sum_{|\bfp|\leq\Lambda_0}E_{|\bfp|}d_{\bfp}\cdot d_{\bfp}^{\dagger}
                   +\sum_{|\bfp|>\Lambda_0}E_{|\bfp|}d_{\bfp}^{\dagger}\cdot d_{\bfp}, \\
  \hat{H}_{DI}
    &\equiv \frac{\omega_{\Lambda_0}^2}{m}
            \sum_{\bfp}\frac{|\bfp|}{\omega_{|\bfp|}}
                       \left(b^{\dagger}_{\bfp}\cdot Md^{\dagger}_{-\bfp}
                             +d_{-\bfp}\cdot M^{\dagger}b_{\bfp}\right).
\end{aligned}
\end{equation}
In (\ref{exchange-interaction-dirac}), $\Lambda_0$ stands for the value of $\Lambda$ where $\rho_D(\Lambda)$ takes a local minimum if it exists. Note that $\hat{H}_{D0}$ is positive definite for the excited states obtained by multiplying creation or annihilation operators on $|\Lambda_0\rangle$.

\begin{figure}
 \includegraphics[scale=0.07]{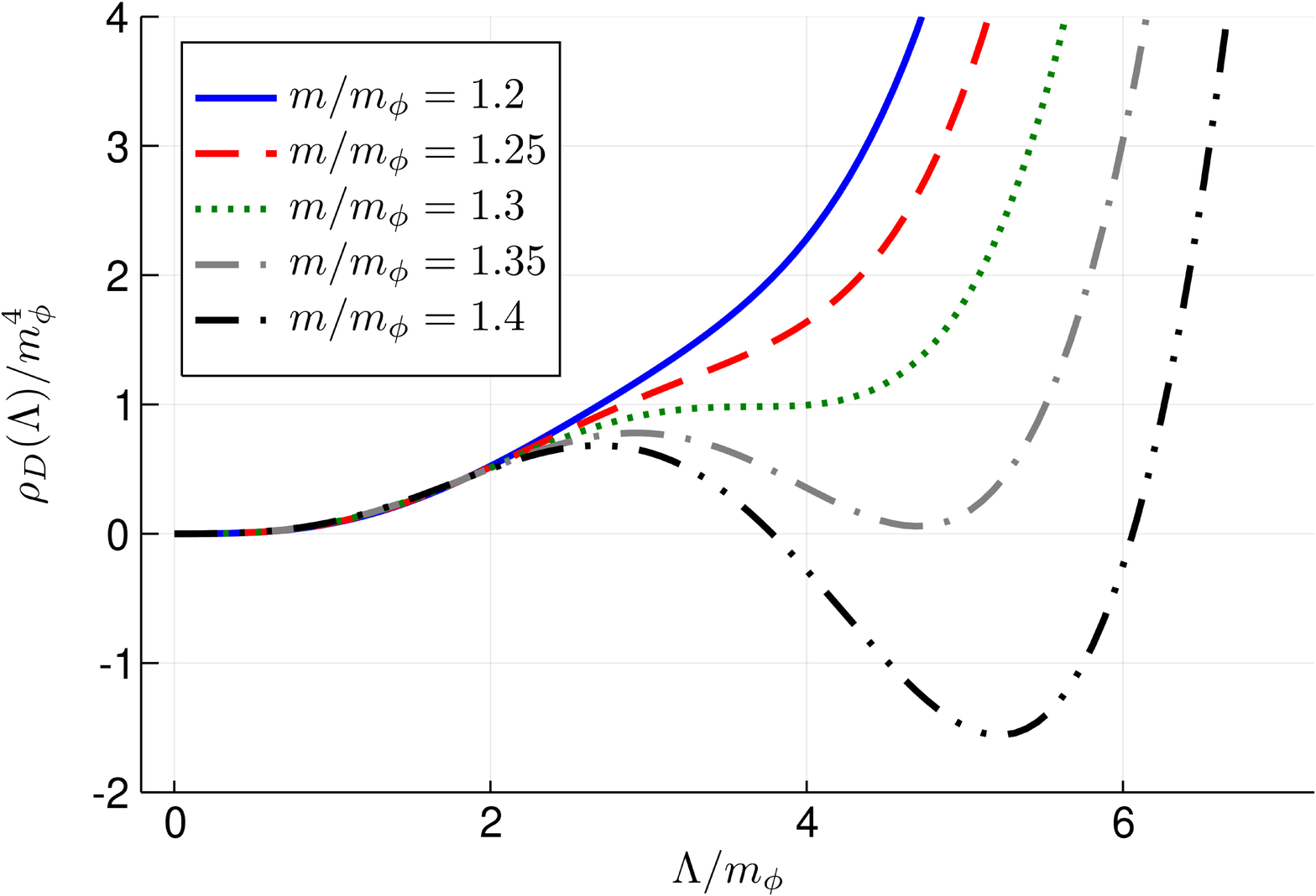}
 \caption{\label{energy-density-dirac}Functional dependence of $\rho_D(\Lambda)$ on $\Lambda\in[0,7m_{\phi}]$ for $g=3$ and $\lambda=0.01$. The horizontal and vertical axis stands for $\Lambda/m_{\phi}$ and $\rho_D(\Lambda)/m_{\phi}^4$, respectively. The graphs for five representative values of $m/m_{\phi}\in[1.2,1.4]$ are shown. A positive local minimum exists at least for $1.3\lesssim m/m_{\phi}\leq 1.35$.}
\end{figure}

When looking for local minima of $\rho_D(\Lambda)$, we consider only the same value of couplings as that in the Majorana case, namely $g=3$ and $\lambda=0.01$. As shown in FIG. \ref{energy-density-dirac}, a positive local minimum of $\rho_D(\Lambda)$ corresponding to the dark energy exists for $1.3\lesssim m/m_{\phi}\leq 1.35$. As we stressed when ${\cal H}'_D$ was defined, the same number of particles and anti-particles are contained in the state $|\Lambda_0\rangle$. However, the pair-annihilation of particles and anti-particles does not occur, because that would increase the energy of the system as a result of the positive definiteness of $\hat{H}_{D0}$. As the Universe expands, particles and anti-particles are produced in pairs via the interaction $\hat{H}_{DI}$, so that the system remains in the metastable state.

\section{conclusion and discussion}
\label{section-conclusion}

In this article, we have proposed a simple model that explains the origin of dark energy with the dynamics of fermion and scalar. For a range of model parameters, a metastable state with positive energy density can exist. This metastable state, or a false vacuum, is stabilized by the attractive and repulsive interactions between fermions mediated by the scalar. \textit{Eternally pair-produced fermions} are also responsible for the stability of the vacuum energy density against the cosmic expansion. We have confirmed, for a certain parameter set, that the metastable dark energy is expected to survive for a period substantially longer than the age of the Universe. The observed value of the vacuum energy density is naturally explained if the fermion has mass of the order $10^{-3}~\textrm{eV}$, which is comparable to the current neutrino mass bound. However, active neutrinos cannot be the candidate for the fermion in the model, because the model does not allow for stable fermionic excitations of the metastable state. Instead, $\psi$ might be a sterile neutrino, the existence of which the LSND anomaly \cite{lsnd} and MiniBooNE experiment \cite{miniboone} have hinted at, though the interpretation of these experimental results is still controversial.

Our model offers a novel scenario for generating the dark energy, but a couple of issues remain to be resolved. First, we have not evaluated the probability of the false vacuum ascending the potential barrier and falling down to the true vacuum through a thermal transition. Unless the temperature of the system is low enough, such a transition is possible. Second, we have used some approximations when deriving the main results in this article. For instance, loop corrections are not taken into account in the effective Lagrangian (\ref{effective-lag}), and a mean field approximation is applied when deriving the Hamiltonian (\ref{hamiltonian-expansion}). It is a non-trivial task to confirm that our results are not qualitatively affected even if we use more accurate calculation techniques. Finally, it will be quite difficult to test our model, because the DE sector must only weakly interacts with the SM sector as postulated at the beginning of section \ref{section-model}. If the above issues are addressed, our model might become another possible solution to the problem of the dark energy.


\end{document}